\documentclass[sigconf,balance=false]{acmart}
\usepackage{popets}

\copyrightyear{YYYY}

\usepackage{multirow}
\usepackage{graphicx}
\usepackage{xcolor}
\usepackage{tcolorbox}
\usepackage{bm}

\newcommand{\defineusercomment}[2]{%
    \expandafter\newcommand\csname #1\endcsname[1]{%
        \begin{tcolorbox}[colframe=white, colback=#2!20, boxrule=0pt, sharp corners]
            \textbf{#1:} ##1 
        \end{tcolorbox}
    }
}

\defineusercomment{kostas}{red}
\defineusercomment{tina}{blue}
\defineusercomment{jiang}{green}
\defineusercomment{dimitris}{purple}

\acmYear{}
\acmVolume{}
\acmNumber{}
\acmDOI{XXXXXXX.XXXXXXX}
\acmISBN{}
\acmConference{}
\begin{document}

\title[]{Preserving Privacy and Utility in LLM-Based Product Recommendations}


\author{Tina Khezresmaeilzadeh}
\affiliation{%
  \institution{University of Southern California}
  \city{Los Angeles}
  \state{California}
  \country{USA}}
\email{khezresm@usc.edu}

\author{Jiang Zhang}
\affiliation{%
  \institution{University of Southern California}
  \city{Los Angeles}
  \state{California}
  \country{USA}}
\email{jiangzha@usc.edu}

\author{Dimitrios Andreadis}
\affiliation{%
  \institution{University of Southern California}
  \city{Los Angeles}
  \state{California}
  \country{USA}}
\email{dgandrea@usc.edu}

\author{Konstantinos Psounis}
\affiliation{%
  \institution{University of Southern California}
  \city{Los Angeles}
  \state{California}
  \country{USA}}
\email{kpsounis@usc.edu}


\renewcommand{\shortauthors}{Khezresmaeilzadeh et al.}


\begin{abstract}
    Large Language Model (LLM)-based recommendation systems leverage powerful language models to generate personalized suggestions by processing user interactions and preferences. Unlike traditional recommendation systems that rely on structured data and collaborative filtering, LLM-based models process textual and contextual information, often using cloud-based infrastructure. This raises privacy concerns, as user data is transmitted to remote servers, increasing the risk of exposure and reducing control over personal information.
    To address this, we propose a hybrid privacy-preserving recommendation framework which separates sensitive from nonsensitive data and only shares the latter with the cloud to harness LLM-powered recommendations. To restore lost recommendations related to obfuscated sensitive data, we design a de-obfuscation module that reconstructs sensitive recommendations locally.
    Experiments on real-world e-commerce datasets show that our framework achieves almost the same recommendation utility with a system which shares all data with an LLM, while preserving privacy to a large extend. Compared to obfuscation-only techniques, our approach improves HR@10 scores and category distribution alignment, offering a better balance between privacy and recommendation quality. Furthermore, our method runs efficiently on consumer-grade hardware, making privacy-aware LLM-based recommendation systems practical for real-world use.
\end{abstract}


\maketitle

\section{Introduction}
Recommendation systems are widely used across e-commerce, streaming platforms, and digital services to deliver personalized content and improve user experience. Traditional approaches, such as collaborative filtering \cite{he2017neural}, content-based methods \cite{_ano_2017}, have been effective but often struggle with challenges like sparse data, cold-start problems, and a limited understanding of user preferences. 
Recent works have demonstrated that applying Machine Learning (ML) models in diverse applications leads to improved accuracy and utility \cite{9299063, 10.1145/2959100.2959190, razmara2024fever, azizi2024efficient, abdollahi2025rocketppa}. Recently, Large Language Models (LLMs) have shown strong capabilities in understanding and generating natural language, making them a promising addition to recommendation systems. By leveraging textual data—such as product descriptions, reviews, and search queries—LLMs offer a way to improve recommendations beyond structured interaction data to even understand user preferences explicitly \cite{he2023large}.
This has motivated researchers to explore their integration into recommendation frameworks, aiming to address the limitations of conventional techniques \cite{xu2025tappingpotentiallargelanguage}.

Despite this promise, there are two primary challenges in deploying LLM-driven recommendation systems in real-world settings. First, As LLMs advance in capability their parameter sizes also grow substantially. State-of-the-art LLMs such as OpenAI’s ChatGPT \cite{OpenAI2025ChatGPT}, Google’s Gemini \cite{Google2025Gemini}, and Anthropic’s Claude \cite{Anthropic2025Claude} often involve billions of parameters, making them computationally demanding and impractical for local deployment on consumer devices. Instead, these models are typically accessed via Application Programming Interfaces (APIs) or proprietary websites. As a result, end users either lack the requisite hardware to deploy such models locally or face vendor restrictions that disallow local downloading and execution. Second, the reliance on remote servers raises critical privacy concerns. Because these models are ``opaque'' in their operation, users cannot fully ascertain how their personal data, such as purchase histories or conversation logs, are processed or stored in black box devices. Moreover, platforms offering access to these models typically require users to sign in, thus linking their activity to identifiable personal accounts and potentially archiving sensitive information. Recent research has also demonstrated that LLMs can infer private attributes about users from their interactions, raising further privacy concerns \cite{staab2023beyond}. For example, purchase histories can uncover sensitive insights about an individual’s health status, financial situation, marital status, or even personal details like skin and hair type, all of which are protected under strict privacy regulations (e.g., HIPAA in healthcare contexts) \cite{HIPAA}.

Recent research has explored privacy-preserving techniques in recommendation systems, see related work section for more details. Non-cryptography-based privacy-preserving techniques, such as differential privacy (DP) \cite{Dwork2006DifferentialP}, anonymization \cite{sweeney2002k} and federated learning \cite{mcmahan2017communication}, have been applied to protect user data.
Additionally, several cryptographic techniques, such as homomorphic encryption (HE) \cite{He2025FedaiFR}, secret sharing \cite{10547416}, and secure multi-party computation (SMPC) \cite{10.1145/3375402}, have been applied to recommendation systems.
However, differentially private language models, while protecting privacy, lose personalization \cite{carranza2023synthetic}. 
Federated learning-based methods, while effective in securing data during training, focus on model training privacy \cite{zhang2024federated, zhao2024llm, zeng2024federated} rather than addressing privacy concerns at the inference stage, which is the main concern when using cloud-hosted LLMs \cite{luo2024privacy}.
Also, cryptographic approaches tend to require significant computational resources and often render data unintelligible, making them incompatible with systems that rely on natural language prompts \cite{cryptoeprint:2023/1147, chen2022x}.
More general, most of these techniques were not developed for recommendation systems that rely on conversation-style text inputs.
Thus, there is a need for privacy-preserving methods that respect the constraints of text-based inference while not sacrificing the recommendation quality.

To address these challenges, we propose a novel hybrid privacy-preserving recommendation framework that balances user privacy and recommendation utility while being able to be run on-device efficiently. Specifically, our framework first employs a context-aware classification model, fine-tuned from BERT \cite{devlin-etal-2019-bert}, to detect which products or user inputs contain sensitive information. As the result of this obfuscation step, only non-sensitive data is offloaded to remote servers for further processing, thus mitigating the privacy risks associated with sharing detailed personal history. Meanwhile, sensitive recommendations are handled locally using a highly compact LLM—Llama 3.2 (1B parameters) \cite{meta2024llama32}. The lightweight Llama 3.2 model can run efficiently on mobile or edge devices, offering multilingual text generation capabilities with minimal latency. By processing these sensitive queries locally, the system avoids transmitting private information to external servers, thereby preserving user privacy and ensuring compliance with legal frameworks \cite{HIPAA}.

Empirical results on real-world e-commerce datasets indicate that our hybrid approach maintains competitive recommendation accuracy (e.g., HR@10) and preserves critical measures like recommendation diversity. Importantly, these performance metrics are achieved alongside meaningful privacy protection, making our method a suitable alternative for personalized recommendation services that need to protect sensitive user information.

In summary, our main contributions are:

\begin{itemize}
    \item An end-to-end privacy-preserving e-commerce recommendation pipeline that separates sensitive and non-sensitive data to protect user privacy while maintaining recommendation quality.
    \item An obfuscator module that utilizes a context-aware product classifier to identify sensitive products, sending only nonsensitive purchase history to the server-based external LLM recommender system.
    \item Utilizing a lightweight local deobfuscator to process sensitive products locally, providing sensitive recommendations.
    \item An assessment of the efficiency of our approach through experiments on standard e-commerce datasets, evaluating privacy leakage, recommendation accuracy (HR@10), and category distribution consistency to compare the diversity of recommendations with the baseline.
\end{itemize}

The rest of the paper is structured as follows:
 Section \ref{sec:problem_statement} defines concepts, goals, and assumptions. Section \ref{sec:methodology} presents our methodology, including the proposed approach and performance metrics. Section \ref{sec:system_design} details the system design, covering the obfuscator, server-based recommendation, deobfuscator, and combined recommendation list. Section \ref{sec:experimental_setup} describes the experimental setup, followed by Section \ref{sec:evaluation}, which evaluates the system. Section \ref{sec:computational_efficiency} discusses findings and Section \ref{sec:limitations} presents limitations, followed by Section \ref{sec:related_work}, which reviews related work on LLM-based recommendations and privacy preservation. Finally, Section \ref{sec:conclusion} concludes and outlines future research directions.

\section{Problem Statement}
\label{sec:problem_statement}

\begin{figure}
    \centering
    \includegraphics[width=0.8\columnwidth]{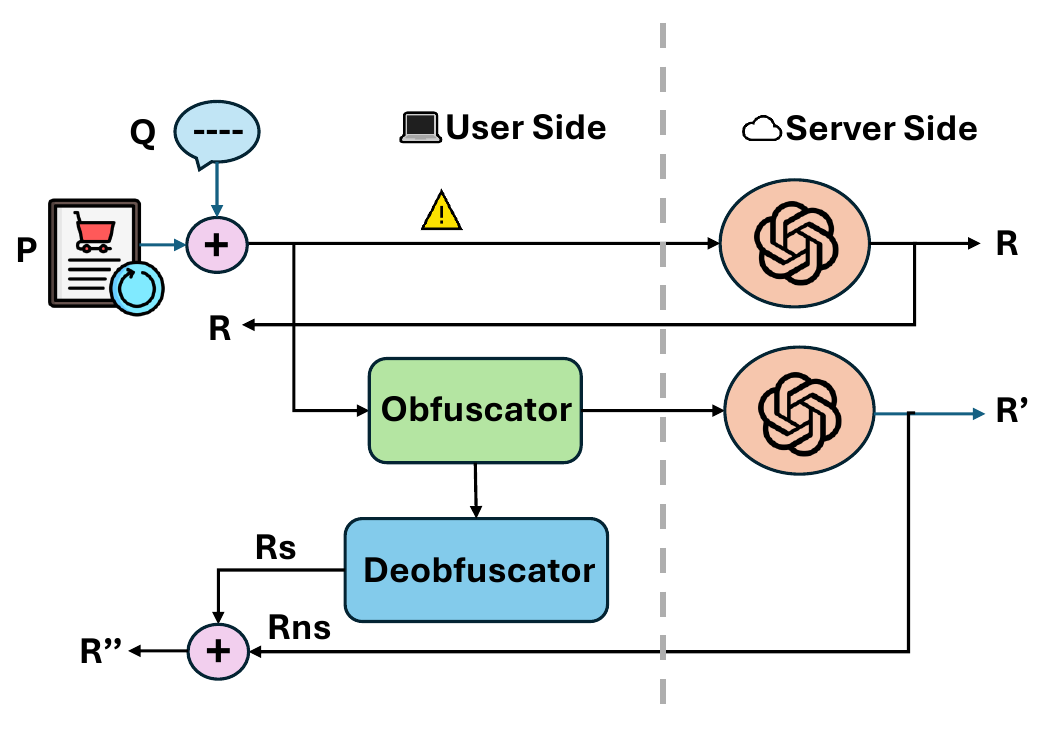}
    \caption{High-level overview of the system.}
    \label{fig:overall-diagram}
\end{figure}


In this section, we define system components (Section \ref{sec:definitions}) and introduce the objectives of this work (Section \ref{sec:goals}) and the threat model (Section \ref{sec:threat-model}). Figure \ref{fig:overall-diagram} provides an overview of the interaction between the user and the server-based recommendation system, demonstrating the communication process and the associated privacy risks.

Recent advancements in large language models (LLMs) have expanded the capabilities of recommendation engines, allowing them to generate context-rich suggestions \cite{10.1145/3678004, wu2024survey, gao2023chat}. However, these models typically run on remote servers, requiring users to send their data, which raises serious privacy concerns \cite{wang2024unique, das2024security}.

A user interacts with the system by submitting a query along with a complete purchase history, which comprises both sensitive and non-sensitive products in order to receive personalized recommendations. Sharing the complete purchase history can lead to privacy leakage, as disclosure of sensitive products enables the server to infer information that the user might prefer to keep private. The objective is to develop a framework that allows users to leverage server-based recommendation systems while preventing the exposure of sensitive product information, without degrading the quality of recommendations.




\subsection{Definitions}
\label{sec:definitions}

\paragraph{Purchase History}
A purchase history is the collection of products previously bought by a user, represented as structured metadata. Each product in the purchase history contains multiple fields, including but not limited to:

\begin{itemize}
    \item \textit{Category:} Broad classification of the product (e.g., "Health \& Personal Care").
    \item \textit{Title:} The product name or short description.
    \item \textit{Features:} Key attributes or specifications.
    \item \textit{Description:} Detailed textual information about the product.
    \item \textit{Details:} Additional metadata or optional information.
\end{itemize}

This structured format captures descriptive and categorical details about the products. It is often utilized in recommendation systems to analyze user preferences and provide personalized suggestions \cite{aws2024metadata}. This is the input of our recommendation system.

\paragraph{Server-based Recommendation System}
Server-based recommendation systems are a type of recommendation system where data processing and recommendation generation are performed on a centralized server. In this work, we focus specifically on systems that leverage large language models (LLMs), accessed via APIs. \\
These systems operate using two primary inputs:
\begin{itemize}
    \item \textit{System Prompt:} A predefined instruction that defines the task for the recommendation system, such as directing the system to generate personalized product recommendations given user purchase history as the input.
    \item \textit{User Prompt:} User-specific data, such as purchase history, that the system utilizes to generate recommendations.
\end{itemize}
The output is a list of $n$ product descriptions that match the user's purchasing patterns and interests. These recommendations aim to provide products that the user may find useful or appealing.

\paragraph{Sensitive Products}
Sensitive products include products in the user’s purchase history that reveal private or personal information. While our framework does not depend on the particular split between sensitive versus non-sensitive products, to make things specific sensitive products are defined as data that may indicate:

\begin{itemize}
    \item \textit{Health Status:} Products designed for individuals with specific health concerns, such as sensitive or dry skin, hair loss, or dietary restrictions.
    \item \textit{Specific Illnesses or Health Conditions:} Products explicitly linked to medical conditions, prescription medications, or medical diagnostic devices.
\end{itemize}

This definition of sensitive data is aligned with privacy principles that prioritize the protection of user health information, which is often considered personally identifiable information (PII) under regulations like HIPAA \cite{HIPAA}. Products classified as sensitive must remain private and be processed locally, as sharing them with a centralized server could expose personal information. This definition is used throughout this work to determine the sensitivity of the products and guide their classification.

\paragraph{Nonsensitive Products} Nonsensitive products are intended for general consumers and do not provide any personally identifiable health information, even if known by others. These products include common consumer goods such as fashion, personal care, household items, electronics, and general wellness products that are not tied to a medical condition. While some nonsensitive products may relate to health or well-being, they are widely used by individuals without specific medical concerns and do not indicate any private health-related information.

\subsection{Goals}
\label{sec:goals}
The proposed system aims to address the trade-off between privacy preservation and recommendation utility in centralized recommendation systems. The key goals are:

1. \textbf{Privacy Preservation:} Sensitive products must remain private and be processed locally on the user’s device.

2. \textbf{Utility:} Despite withholding sensitive data, the system must deliver high-quality recommendations, including recommendations related to sensitive products, evaluated through:
\begin{itemize}
    \item \textit{Prediction Accuracy (Hit Rate)}, which measures how well the recommendations align with user preferences.
    \item \textit{Category Distribution Alignment}, which measures how well the distribution of product categories in the recommendations matches that of a baseline system using the full purchase history.
\end{itemize}
To address these challenges, the system must first distinguish between sensitive and nonsensitive products in the user’s purchase history. Nonsensitive products can be processed externally by a powerful server-based large language model (LLM) to leverage its advanced contextual capabilities, while sensitive products are handled locally to preserve privacy.

The ultimate goal is to deliver a unified recommendation list that balances privacy, utility, and quality, ensuring compliance with privacy constraints without sacrificing the user experience.

\subsection{Threat Model}
\label{sec:threat-model}


 As illustrated in Figure~\ref{fig:overall-diagram}, the obfuscator identifies sensitive products in the purchase history of the user and removes them before sending the data to the server-based recommendation system. In parallel, the deobfuscator processes these sensitive products locally, generates relevant sensitive recommendations, and merges them with the nonsensitive recommendations returned by the server. This approach enables personalized suggestions while keeping the sensitive products list private.


\paragraph{User} The user’s objective is to share a purchase history with the server-based recommendation system and receive personalized product suggestions without revealing sensitive products to the server. To achieve this, the user employs the local obfuscation-deobfuscation framework mentioned above. This configuration restricts the server’s access to private data by confining sensitive products processing to the user’s device. The threat model assumes that the user’s device is trusted, secure and not compromised by malware or adversarial manipulation.

\paragraph{Server-based Recommendation System} The goal of the server-based recommendation system is to generate product suggestions based on the user’s provided purchase history. Unlike real-world e-commerce platforms such as Amazon.com \cite{amazon} or Google Shopping \cite{google-shopping}, which retrieve product information from live databases, this system does not have real-time access to product listings, inventory, or user transaction logs. Instead, it relies on pre-trained data and provides recommendations based on learned knowledge rather than platform-specific product suggestions. \\
The system processes only the purchase history explicitly included in each input prompt. Unlike platforms such as Amazon.com, where transaction records are permanently logged and cannot be removed by the user \cite{amazon-privacy-policy}, we assume a recommendation system where the system allows users to exclude specific purchase data before sending it to the server, limiting what information is exposed. \\
The server operates under a semi-honest model, meaning it follows the defined protocol but may attempt to infer user personal data from the received input prompt. However, it does not have access to external sources, cross-site tracking mechanisms, or any data stored locally on the user’s device. The server processes only the data explicitly provided in each prompt and does not supplement it with external sources. Furthermore, the system has sufficient computational resources to handle recommendation generation efficiently.

\section{Methodology}
\label{sec:methodology}


\subsection{Proposed Method}
\label{sec:proposed_method}
The system processes a user's purchase history to generate personalized recommendations while preserving privacy.
Let $\mathcal{P} = \{p_1, p_2, \dots, p_m\}$ denote a user's complete purchase history, where each product $p_i$ is associated with textual metadata including its main category, title, features, and description. The system splits $\mathcal{P}$ into two disjoint subsets, $P_s = \{\text{Sensitive products}\}$ and $P_{ns} = \{\text{Nonsensitive products}\}$:

\begin{itemize}
    \item $P_s$: Products whose metadata potentially reveals private or sensitive information (e.g., specific health conditions).
    \item $P_{ns}$: The subset of products for general users considered safe for external processing.
\end{itemize}
The sizes of these sets are denoted as $|P_s|$ and $|P_{ns}|$, representing the number of sensitive and nonsensitive products, respectively. \\
Based on this partition, the system generates two categories of recommendations:

\begin{itemize}
    \item $R_s$: Recommendations derived from $P_s$ (sensitive products), generated locally on the user's device to preserve privacy.
    \item $R_{ns}$: Recommendations derived from $P_{ns}$ (nonsensitive products), generated using a server-based recommendation system.
\end{itemize}

Let $n_s$ and $n_{ns}$ represent the number of recommendations derived from $P_s$ and $P_{ns}$ respectively, that is, 
$n_s=|R_s|$ and
$n_{ns}=|R_{ns}|$.
The final recommendation list $R''$ is produced by merging $R_s$ and $R_{ns}$ and it is of size $n=n_s+n_{ns}$.

\begin{figure*}[]
    \centering
    \includegraphics[width=\textwidth]{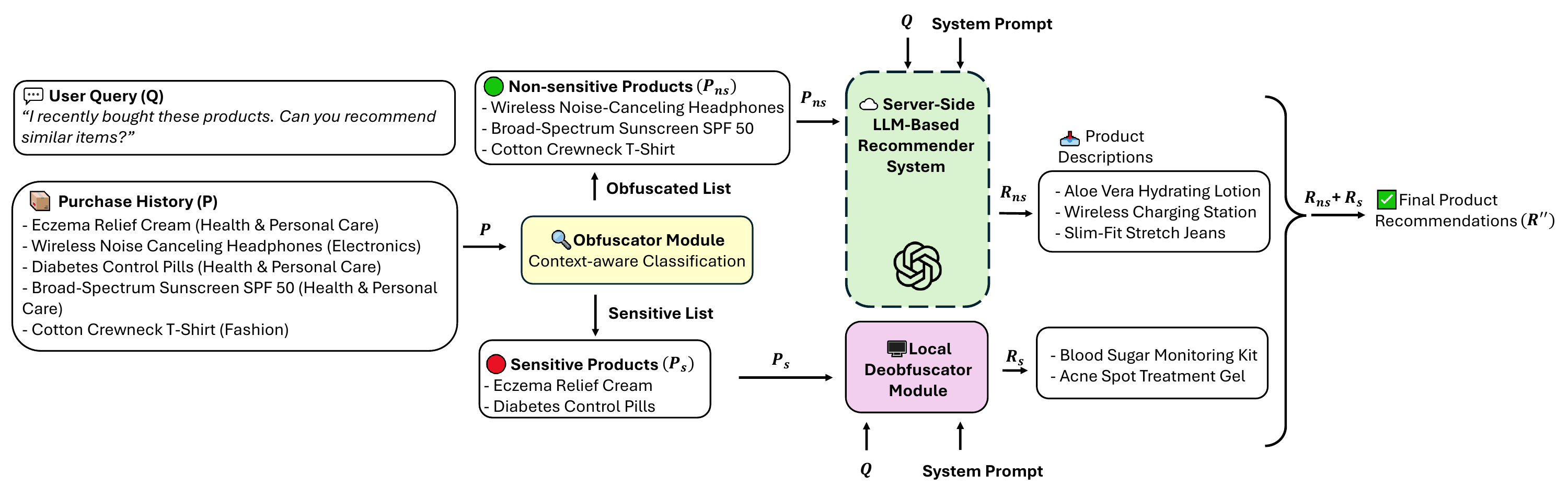}
    \caption{Overview of different components of privacy-preserving recommendation system. The user query ($Q$) and purchase history ($\mathcal{P}$) are the inputs to the system. $\mathcal{P}$ is processed by the obfuscator, which classifies products into sensitive ($P_s$) and nonsensitive ($P_{ns}$) categories.  $Q$ and $P_{ns}$ are sent to the server-side LLM recommender, while $P_s$ is processed locally using deobfuscator module.  The final recommendation set is obtained by combining the locally processed sensitive recommendations ($R_s$) with the server-generated nonsensitive recommendations $R_{ns}$.
    }
    \label{fig:overview-of-the-pipeline}
\end{figure*}

\medskip
Figure \ref{fig:overview-of-the-pipeline} illustrates an overview of the entire architecture, which comprises the following components:

\paragraph{Obfuscator:} Processes the user's purchase history to separate sensitive ($P_s$) and nonsensitive ($P_{ns}$) products. This classification is handled using a fine-tuned BERT model that employs a weighted focal loss to prioritize the detection of sensitive information while reducing false negatives. The model relies on textual metadata to make predictions. 

\paragraph{Server-Based Recommendation System:} The nonsensitive product subset ($P_{ns}$) is processed by a powerful server-based LLM to generate nonsensitive recommendations ($R_{ns}$). The model uses a zero-shot system prompt alongside user query ($Q$) and product metadata from $P_{ns}$ as the user prompt to produce product recommendations. Only nonsensitive data is sent to the server, keeping private information on the user's device. 

\paragraph{Local Deobfuscator:} The sensitive product subset ($P_{s}$) is processed locally on the user's device using a lightweight local LLM to generate sensitive product recommendations ($R_s$).
Note that the final recommendation list ($R''$) is constructed by combining $R_{ns}$ and $R_s$ after resolving conflicts and ranking adjustments.

\medskip
Looking at Figure \ref{fig:overview-of-the-pipeline}, the overall data flow is as follows: the obfuscator classifies products as either sensitive ($P_{s}$) or nonsensitive ($P_{ns}$), with $P_{ns}$ being sent to the server-based recommendation system and $P_{s}$ processed locally by the deobfuscator, before merging the respective recommendations ($R_{ns}$, $R_s$) into the final recommendation set ($R''$).



To assess the recommendation quality, we use a local vector-based retriever built with ChromaDB \cite{chromadb}. Each generated recommendation is compared to an indexed product database to retrieve the most semantically similar product, allowing us to evaluate how well recommendations align with actual product options.

\subsection{Performance Metrics}
\label{sec:performance_metrics}
To assess the system's performance comprehensively, we evaluate both utility (the quality and relevance of recommendations) and privacy preservation (the impact of obfuscation on sensitive data removal). Our evaluation uses the following utility and privacy metrics:

\subsubsection{Utility}


HR@10 is a widely recognized metric in recommendation systems, used to evaluate the relevance and accuracy of recommendations. It measures whether the system successfully includes the actual target product within the top-10 recommended items. In our system, HR@10 assesses whether the recommendation system retrieves the $(n+1)_{th}$ product in a user's purchase sequence, given the first $n$ products in their purchase history.

However, due to the scale and diversity of our dataset (see Section \ref{sec:dataset}), HR@10 is often not informative, as the probability of retrieving the exact target product is extremely low. To address this, we introduce two alternative metrics: \emph{Category-Based HR@10} and \emph{Semantic Similarity HR@10}, which better capture the relevance of recommendations in a practical setting.


\textbf{1. Category-Based HR@10:}
Category-Based HR@10 measures the system's ability to recommend items within the same main category as the target product. A hit is recorded if the main category of the target product matches the main category of any of the top-10 recommendations.
This approach ensures that the system captures category-level relevance, reflecting its ability to suggest products within the same domain as the user's historical preferences. The main categories are derived from product metadata, and matching is performed using a vector-based database retriever that maps recommended product descriptions to their respective categories.

\textbf{2. Semantic Similarity HR@10:}
To evaluate the system's ability to recommend conceptually related products, we compute the cosine similarity between the target product's ($n+1$) description and each of the top-10 recommended product descriptions \cite{cosine-similarity}. Cosine similarity measures the alignment between two vector embeddings in a high-dimensional space and is calculated as:
\begin{equation}
\text{Cosine Similarity} = \frac{\mathbf{u} \cdot \mathbf{v}}{\|\mathbf{u}\| \|\mathbf{v}\|},
\end{equation}
where $u$ and $v$ are the embeddings of the target product and a recommended product, respectively. A similarity score of 1 indicates perfect semantic alignment, while 0 implies no similarity.

Product descriptions are transformed into embeddings using a pre-trained Sentence Transformer model, such as ``all-MiniLM-L6-v2,'' \cite{all-MiniLM-L6-v2} which maps text into a 384-dimensional dense vector space optimized for clustering and semantic search. These embeddings are used in the cosine similarity calculations to quantify the contextual and semantic alignment between the target product and the recommendations.

For each user's purchase history, the maximum cosine similarity among the top-10 recommendations is recorded as the semantic similarity HR@10. 

Note that Semantic Similarity HR@10 complements Category-Based HR@10 by providing an additional perspective on the system's recommendation accuracy. While category matching captures broader relevance by aligning recommendations within the same category, Semantic Similarity HR@10 evaluates the relationships between products based on their contextual and semantic meaning. This ensures that the system can identify meaningful connections even when products are categorized differently or vary significantly within the same category.

\textbf{Category Distribution Metrics.}
Preserving the diversity of recommendations is critical for ensuring that a privacy-preserving recommendation system continues to reflect user preferences without skewing toward nonsensitive categories. To evaluate this, we use category distribution vectors and compute deviations from the baseline system using L1 and L2 distances. These metrics provide a robust framework to assess system-wide diversity and balance across product categories.

For a recommendation list of $n$ items and $C$ categories, the category distribution vector is defined as 
$v = [v_1, v_2,..., v_C],$
where $v_i$ represents the proportion of recommendations belonging to category $i$, that is, $v_i = \frac{\text{count of items in category } i}{\; n \;}$, for $i=1\ldots C$, and 
$ \sum_{i=1}^C v_i = 1$.

\paragraph{Distance Metrics for Evaluation}
To evaluate how well the obfuscated systems replicate the category diversity of the baseline, we compute the L1 distance and L2 distance between their respective category distribution vectors.

\textbf{L1 Distance:} The L1 distance measures the total absolute difference between the baseline ($v_{baseline}$) and obfuscated system ($v_{system}$) distributions:

\begin{equation}
L_1 = \sum_{i=1}^C |v_{\text{baseline},i} - v_{\text{system},i}|.
\end{equation}
This metric measures overall imbalances in category distributions. A smaller L1 distance indicates better preservation of category balance, reflecting minimal disruption to the original diversity of recommendations.

\textbf{L2 Distance:} The L2 distance, or Euclidean distance, quantifies the magnitude of deviations:
\begin{equation}
L_2 = \sqrt{\sum_{i=1}^C (v_{\text{baseline},i} - v_{\text{system},i})^2}.
\end{equation}

The L2 distance emphasizes larger deviations, making it sensitive to noticable changes in category proportions. A smaller L2 value indicates a closer alignment with the baseline distribution.

Users expect recommendations to capture their diverse preferences. However, noticable deviations in category distributions can result in imbalanced recommendations, either by overrepresenting non-sensitive categories or omitting sensitive ones. By using L1 and L2 distances, we can measure these deviations to ensure that obfuscation and deobfuscation mechanisms maintain the diversity present in a user's purchase history.
Unlike HR@10, which focuses on individual-level recommendation relevance, category distribution metrics provide a system-wide perspective. They evaluate how obfuscation and deobfuscation affect the overall balance of recommendations across all categories,  addressing both the accuracy for individual users and fairness across the system. 
The HR@10 and distance metrics together capture both the relevance of recommendations for individual users and the preservation of category diversity across the system.

\subsubsection{Privacy}

We measure privacy exposure using two metrics: \emph{Binary Privacy Leakage} ($\text{PL}_b$) and \emph{Semantic Privacy Leakage} ($\text{PL}_s$). The first metric quantifies the fraction of leaked sensitive products, while the second accounts for their varying sensitivity levels.

\paragraph{Binary Privacy Leakage ($\text{PL}_b$)}
$\text{PL}_b$ measures the average fraction of sensitive products that have been leaked. It is defined as:

\begin{equation}
    \text{PL}_b = \frac{1}{n_s} \sum_{i=1}^{n_s} x_i,
\end{equation}
where $n_s$ is the total number of sensitive products in the purchase history, and $x_i = 1$ if the $i$th sensitive product is leaked and 0 otherwise.

\paragraph{Semantic Privacy Leakage ($\text{PL}_s$)}
$\text{PL}_s$ extends $\text{PL}_b$ by incorporating sensitivity scores of the products, assigning higher importance to more sensitive items. It is defined as:

\begin{equation}
    \text{PL}_s = \frac{\sum_{i=1}^{n_s} x_i s_i}{\sum_{i=1}^{n_s} s_i},
\end{equation}
where $s_i$ represents the sensitivity score of the $i$th sensitive product. The numerator accumulates the sensitivity scores of leaked products, while the denominator normalizes by the total sensitivity of all sensitive products.

Both metrics assume that sensitivity scores are predefined and reflect the relative importance of different products. If sensitivity scores are not available, $\text{PL}_b$ serves as a simpler alternative. However, $\text{PL}_s$ provides a more refined measure by accounting for varying degrees of sensitivity among leaked products.

To measure sensitivity scores, we use few-shot prompting with ChatGPT-4-o \cite{OpenAI2025ChatGPT4} to assign a sensitivity score between 0 and 1 to each product based on its perceived sensitivity. The exact few-shot prompt used for this evaluation is provided in Appendix \ref{sec:prompt-for-sensitivity-score-assignment}. For the products used in our analysis (see Section \ref{sec:utility_and_privacy_results}), the mean sensitivity of the sensitive products (i.e., those labeled with a ground-truth label as sensitive) is 0.5422, and the standard deviation is 0.2317.


\section{System Design}
\label{sec:system_design}
\subsection{Obfuscator}
\label{sec:obfuscator}
The Obfuscator module enhances user privacy by separating products in the purchase history ($\mathcal{P}$) into sensitive ($P_s$) and nonsensitive ($P_{ns}$) subsets. Sensitive products ($P_s$) may reveal private information about the user (e.g., health-related conditions), while nonsensitive products ($P_{ns}$) are safe for external processing. This classification is achieved through two methods: the Categorical Obfuscator and the BERT-based Obfuscator.

\subsubsection{Data Preparation}
\label{sec:data_preparation}
Each product in $\mathcal{P}$ is represented as a textual string derived from the following fields:

\begin{itemize}
    \item \textit{Main Category:} Broad product categorization (e.g., ``Health \& Household,'' ``Beauty \& Personal Care'').
    \item \textit{Title:} The product’s short name or headline.
    \item \textit{Features:} Key attributes of the product, often presented as bullet points.
    \item \textit{Description:} Detailed text providing an overview of the product.
    \item \textit{Details:} Supplementary metadata, when available.
\end{itemize}
This representation presents the input for the obfuscation methods, enabling a complete view of the product's attributes and context. While this detailed representation is used for the BERT-based approach, the Categorical Obfuscator relies only on predefined category mappings.

\subsubsection{Categorical Obfuscator}
\label{sec:categorical_obfuscator}
This method involves filtering out products belonging to predefined sensitive categories (e.g., Health \& Personal Care, Health \& Household, and Beauty \& Personal Care) from the user's purchase history. By utilizing a fixed mapping of categories to sensitivity, the approach assumes that all products within these categories are inherently sensitive. While this method is computationally efficient and straightforward—requiring no model training—it presents limitations. False positives may occur when nonsensitive products (e.g., general-use beauty items) are misclassified as sensitive, and false negatives may arise when sensitive products outside the predefined categories (e.g., books on health conditions) are overlooked.

\subsubsection{BERT-based Obfuscator}
\label{sec:BERT_based_obfuscator}
To address the limitations of the categorical obfuscator, we leverage BERT (Bidirectional Encoder Representations from Transformers) \cite{devlin-etal-2019-bert}, a state-of-the-art transformer-based model, fine-tuned specifically on sensitive product categories, to classify products based on their full textual descriptions rather than relying solely on predefined categories. Additionally, BERT's pretrained architecture allows for efficient fine-tuning on domain-specific datasets, making it well-suited for this task \cite{sun2019fine}.

By employing context-aware classification, nonsensitive products that fall under a broad category will not be misclassified as sensitive. This reduces the difference between a user's actual purchase history and their obfuscated purchase history. As a result, the obfuscated profile remains closer to the original, making the obfuscation process more precise and effective.
Further details on the training process and dataset construction for the BERT-based obfuscator are provided in Section \ref{sec:training_and_testing_details}.

\textit{Inference and Classification:} During inference, product descriptions are processed through the fine-tuned BERT model to produce sensitivity scores which are utilized as follows:

\begin{itemize}
    \item Scores > 0.5: Classified as $P_s$(sensitive).
    \item Scores $\leq$ 0.5: Classified as $P_{ns}$(nonsensitive).
\end{itemize}

\subsection{Server-Based Recommendation System}
The Server-Based Recommendation System generates recommendations for the nonsensitive subset ($P_{ns}$) by leveraging a cloud-based large language model (LLM), ensuring privacy while maximizing utility. By exclusively processing nonsensitive data, the system avoids any potential leakage of sensitive information.

\subsubsection{Model Choice}
\label{sec:model_choice}
We employed ChatGPT-4-o \cite{OpenAI2025ChatGPT4} via its API for its state-of-the-art performance in generating contextually rich and relevant product recommendations \cite{di2023evaluating}. Its ability to balance semantic understanding and categorical diversity makes it particularly well-suited for this task. Preliminary evaluations showed that ChatGPT-4-o performs well in terms of recommendation relevance and diversity, making it an appropriate choice for this application.

\subsubsection{Prompt Design and Validation}
\label{sec:prompt_design_and_validation}
To ensure that the recommendations align with the user’s purchase history, the system employs a carefully crafted prompt. This prompt ensures a proportional representation of categories and comprehensive coverage across all categories in $P_{ns}$.

The number of nonsensitive recommendations ($n_{ns}$) is calculated as:
\begin{equation}
    n_{ns} = n \times \frac{|P_{ns}|}{|P_{ns}| + |P_{s}|},
\end{equation}
where $n$ is the total number of desired recommendations (set to 10 in the experiments), $|P_{ns}|$ is the number of nonsensitive products in the purchase history, and $|P_{s}|$ is the number of sensitive products in the purchase history. The full prompt used in the server-based recommendation system is provided in Appendix \ref{sec:prompt-for-server-based-recommendation-system}.

\subsection{Deobfuscator}
\label{sec:deobfuscator}
The Deobfuscator complements the server-based recommendation system by generating recommendations for the sensitive products in $P_s$. This step is performed locally on the user’s device to maintain strict privacy standards, ensuring that sensitive data is never transmitted externally.
 
For sensitive products ($P_s$), their concatenated textual representations are processed locally using Llama 3.2 1B \cite{Llama-3.2-1B}, a lightweight language model capable of efficient inference on consumer-grade hardware. By handling sensitive recommendations locally, the system ensures that privacy-critical information remains on the user’s device.

\subsubsection{Prompt Design}
\label{sec:prompt_design}
The same prompt structure used for nonsensitive recommendations is adapted for sensitive products, see  Appendix \ref{sec:prompt-for-local-deobfuscator} for the full prompt. 

\subsubsection{Output:}
The model generates the sensitive recommendation list ($R_s$) locally. These recommendations are stored securely on the user’s device, ensuring strict privacy standards. 
The number of sensitive recommendations ($n_s$) is calculated as:
\begin{equation}
    n_{s} = n \times \frac{|P_{s}|}{|P_{ns}| + |P_{s}|},
\end{equation}
This ensures that sensitive recommendations are proportional to the sensitive product share in the purchase history. By processing sensitive data locally, the system safeguards privacy while still providing personalized recommendations.

\subsection{Combined Recommendation List} 
\label{sec:combined_recommendation_list}
The final recommendation list integrates both nonsensitive and sensitive recommendations, providing the final set of personalized suggestions:
\begin{equation}
    R'' = R_{ns} + R_{s},
\end{equation}
where $R''$ represents the complete recommendation list, including both non-sensitive and sensitive product recommendations. By combining the outputs from the server and the local model, the system delivers personalized recommendations without compromising the user's privacy.

\section{Experimental Setup}
\label{sec:experimental_setup}
To thoroughly evaluate our privacy-preserving recommendation system we consider three setups: A baseline system with no obfuscation, a system that uses either the Categorical Obfuscator or the BERT-based Obfuscator, and a system that uses either of the obfuscators and the Llama-based Deobfuscator. This experiments are designed to quantify the trade-offs between privacy preservation and utility recovery, providing insights into how effectively the system balances these two objectives.

\subsection{Dataset}
\label{sec:dataset}
\begin{table}[]
\begin{tabular}{lll}
\hline
\textbf{Category}                    & \textbf{\#Users} & \textbf{\#Items} \\ \hline
Amazon\_Fashion             & 2.0M    & 825.9K  \\ \hline
Beauty\_and\_Personal\_Care & 11.3M   & 1.0M    \\ \hline
Electronics                 & 18.3M   & 1.6M    \\ \hline
Health\_and\_Household      & 12.5M   & 797.4K  \\ \hline
Magazine\_Subscriptions     & 60.1K   & 3.4K    \\ \hline
Books                       & 10.3M   & 4.4M    \\ \hline
Baby\_Products              & 3.4M    & 217.7K  \\ \hline
Grocery\_and\_Gourmet\_Food & 7.0M    & 603.2K  \\ \hline
Health\_and\_Personal\_Care & 461.7K  & 60.3K   \\ \hline
Musical\_Instruments        & 1.8M    & 213.6K  \\ \hline
\end{tabular}
\caption{ Summary of the selected categories from the Amazon Reviews 2023 dataset, including the number of users and items across sensitive and nonsensitive domains.}
\label{tab:category-distribution}
\end{table}

The system leverages the 2023 Amazon Reviews Dataset McAuley Lab \cite{amazon-dataset2023}, a large-scale dataset that includes product reviews and metadata across a variety of categories. This dataset provides a rich resource for evaluating the system’s ability to handle both recommendation and privacy-preserving tasks. The selected categories reflect a mix of sensitive and nonsensitive domains, ensuring a diverse range of purchasing behaviors is represented.

To evaluate the system’s performance comprehensively, ten categories were selected from the dataset, comprising both sensitive and nonsensitive domains. Sensitive categories—``Health \& Personal Care,'' ``Health \& Household,'' and ``Beauty \& Personal Care''—were deliberately chosen for their relevance to privacy concerns. Nonsensitive categories were included to capture a broader diversity of textual characteristics and purchasing contexts, essential for robust system evaluation. Table \ref{tab:category-distribution} summarizes the selected categories.

\paragraph{Classification of Sensitive and Nonsensitive Categories:}
Sensitive categories were selected based on their strong potential to reveal private user information. Specifically:

\begin{itemize}
    \item ``Health \& Personal Care'' and ``Health \& Household'' often include products that relate to medical conditions or personal wellness.
    \item ``Beauty \& Personal Care'' contains items designed for individuals with specific health concerns, such as sensitive or dry skin or hair, etc.
\end{itemize}

While it is possible that certain nonsensitive categories could occasionally contain products that may reveal private information depending on how privacy is defined, as already discussed in Section \ref{sec:definitions}, we use health, personal care, and body-related items as sensitive since this aligns with common privacy concerns in e-commerce contexts, where such categories are more likely to reveal intimate or personal details about users. Consequently, other categories are treated as nonsensitive.


To construct user purchase histories, we first  filter the dataset to "engaged'' users (users having at least 30 products in their purchase history), then we sorted each user's products in chronological order and selected the last 20 purchases. This approach maintains a consistent-length input across all users while prioritizing the most recent purchases, which better capture current user preferences and behavior. Choosing 20 purchases provides a balance between computational efficiency and retaining sufficient context to generate high-quality recommendations.


\subsection{Experimental Setups}
\label{sec:experimental_baselines}
For both obfuscation methods we define three experimental setups (as demonstrated in Figure \ref{fig:overall-diagram}) to systematically evaluate the system:

\textbf{Baseline ($R$):}
The complete purchase history, including sensitive products, is sent to the server-based recommendation system. This serves as the upper bound for recommendation quality (utility) since no privacy-preserving mechanisms are applied. The Baseline provides a reference point to measure the impact of obfuscation and deobfuscation on utility.

\textbf{Only Obfuscator ($R'$):}
Under this setup, sensitive products are removed from the purchase history using either the Categorical Obfuscator or the BERT-based Obfuscator. The remaining nonsensitive products are then sent to the server-based recommendation system. This condition highlights the privacy-preserving aspect of the system but sacrifices utility, as sensitive products are completely excluded from the recommendations.

\textbf{Obfuscator + Deobfuscator ($R''$) :}
This setup combines the obfuscator with the deobfuscator to address the utility loss introduced by removing sensitive products from the shared purchase history. Nonsensitive recommendations are generated by the server-based system using the obfuscated purchase history. Simultaneously, the deobfuscator generates recommendations for sensitive products locally using Llama 3.2 1B. The final recommendation list merges nonsensitive and sensitive recommendations. This setup aims to preserve privacy while recovering utility, making it the most balanced option.

\subsection{Obfuscator Training and Testing Details}
\label{sec:training_and_testing_details}

As already discussed, we use a finetuned BERT-based model for the obfuscation of the sensitive products. The model training details are given below.

\paragraph{Dataset Construction} 
We randomly sampled $10,000$ product entries from predefined sensitive categories to create our dataset. Each product entry comprised a title, its main category, and a list of features. We concatenated these attributes into a single textual prompt to serve as the input to our model. An example prompt format (without showing actual code) included the product’s title on the first line, followed by main category information, and then a list of features.

\paragraph{Labeling} Products were labeled as either ``sensitive'' or ``nonsensitive'' following the definition of sensitive and nonsensitive products outlined in Section \ref{sec:definitions}.
ChatGPT-4o was used to assign these labels based on the classification criteria. The exact few-shot prompt used for this classification is provided in appendix \ref{sec:prompt-for-product-sensitivity-evaluation}. The model follows specific instructions for the classification and makes uses of few examples so as to better infer the correct labeling pattern.

\paragraph{Model and Tokenization} We fine-tuned a BERT-based model (\textit{bert-base-uncased}) for binary classification of product sensitivity. Each product description was tokenized using the BERT tokenizer with a maximum length of 256, applying right padding and truncation to keep input sizes consistent.

\begin{table}[]
\begin{tabular}{cc}
\hline
\textbf{Hyperparameter} & \textbf{Value}     \\ \hline
Learning Rate           & $2 \times 10^{-5}$ \\ \hline
Weight Decay            & 0.01               \\ \hline
Batch Size              & 16                 \\ \hline
Epochs                  & 5                  \\ \hline
\end{tabular}
\caption{Hyperparameters used for training the BERT-based product sensitivity classification model.}
\label{tab:BERT-parameters}
\end{table}
\paragraph{Training and Hyperparameters} 
The dataset was split into training (70\%), validation(20\%), and test (10\%) subsets using stratified sampling to maintain class balance in training, evaluation, and test set. Three actions were taken to improve detection of sensitive products, as misclassifying a sensitive product as nonsensitive is more critical than the reverse.

First, we employ Focal Loss \cite{lin2017focal} to modify the standard cross-entropy loss. This loss down-weights well-represented examples and concentrates learning on less-represented examples which are harder to classify. The loss function is defined as:

\begin{equation}
\label{eq:loss-function}
    L = - \sum_{i=1}^{N} wc_i (1 - pc_i)^\gamma \log(pc_i),
\end{equation}
where $N$ is the total number of samples, $wc_i$ is the weight of the true class of data sample $i$, and $pc_i$ is the predicted probability of the true class of data sample $i$. The focusing parameter $\gamma$ is set to 2 in this implementation. The class weight $wc_i$ is given by:
\begin{equation} wc_i = \frac{N}{2 \cdot N_c},\end{equation}
where $N_c$ is the number of samples in class $c$ where data sample $i$ belongs to. This approach increases the loss contribution of the (``sensitive'') labeled products as they tend to be less represented.

Second, we lower the decision threshold for the ``sensitive'' class to 0.3. This change increases recall by classifying more products as sensitive, thereby reducing the risk of mistakenly labeling truly sensitive items as nonsensitive.

Finally, model selection is based on the best F1-score on the validation set. This metric optimizes the trade-off between precision and recall, aligning with our goal of accurately identifying sensitive products while minimizing misclassification.

Table \ref{tab:BERT-parameters} summarizes the  hyperparameters used during the training process. 
During each training epoch, the model was optimized to minimize the loss mentioned in Eq. \ref{eq:loss-function}. A checkpoint was saved at the end of every epoch, and the best-performing checkpoint on the validation set (based on F1 score) was automatically loaded at the end of training.

\paragraph{Evaluation Metrics}
Model performance was evaluated using four standard classification metrics: accuracy, precision, recall, and F1-score. The final model achieved an evaluation loss of 0.181, an accuracy of 0.911, an F1-score of 0.868, a precision of 0.815, and a recall of 0.928. These results indicate that the model effectively differentiates between sensitive and nonsensitive products while maintaining a trade-off between capturing sensitive items and avoiding unnecessary misclassification.

\begin{table}[]
\begin{tabular}{ll}
\hline
\textbf{Component}  & \textbf{Specification}                           \\ \hline
\textbf{GPU}        & 2× NVIDIA RTX A6000 (49GB VRAM)             \\ \hline
\textbf{CPU}        & AMD EPYC 7662 (64 Cores, 128 Threads)  \\ \hline
\textbf{RAM}        & 128GB DDR4                                       \\ \hline
\textbf{Storage}    & 1TB NVMe SSD  \\ \hline
\textbf{Networking} & 1 Gbps Ethernet (for API calls to global LLM)    \\ \hline
\end{tabular}
\caption{Hardware Configurations of the System}
\label{tab:hardware-configuration}
\end{table}
\paragraph{Testbed Configuration}
Table~\ref{tab:hardware-configuration} presents the hardware configuration of the testbed used for our experiments. We evaluate system hardware usage and end-to-end latency 
in Section~\ref{sec:computational_efficiency}.

\section{Evaluation}
\label{sec:evaluation}

\begin{table*}[]
\begin{tabular}{lllllll}
\hline
\multicolumn{1}{c}{\multirow{2}{*}{\textbf{Scheme}}} & \multicolumn{2}{c}{\textbf{HR@10}}       & \multicolumn{2}{c}{\textbf{Distance}}       & \multicolumn{2}{c}{\textbf{Privacy Leakage (\%)}} \\ \cline{2-7} 
\multicolumn{1}{c}{}                                 & \textbf{Categorical} & \textbf{Semantic} & \textbf{L2 Distance} & \textbf{L1 Distance} & \textbf{$\bm{PL_b}$ (\%)}       & \textbf{$\bm{PL_s}$ (\%)}       \\ \hline
Baseline ($R$)                                         & 0.6263               & 0.3045            & 0                    & 0                    & 100\%                   & 100\%                   \\ \hline
Only Local                                           & 0.4667               & 0.2398            & 0.4538               & 0.7711               & \textbf{0\%}            & \textbf{0\%}            \\ \hline
Categorical Obf Only ($R$)                            & 0.4742               & 0.2444            & 0.4860               & 0.8462               & \textbf{0\%}            & \textbf{0\%}            \\ \hline
Categorical Obf + Deobf ($R''$)                        & 0.5258               & 0.2994            & 0.2847               & 0.4773               & \textbf{0\%}            & \textbf{0\%}            \\ \hline
BERT Obf Only ($R'$)                                   & 0.5960               & 0.2866            & 0.2672               & 0.4610               & 22.2833\%               & 10.9899\%               \\ \hline
BERT Obf + Deobf ($R''$)                               & \textbf{0.6061}      & \textbf{0.3117}   & \textbf{0.2267}      & \textbf{0.3747}      & 22.2833\%               & 10.9899\%               \\ \hline
\end{tabular}
\caption{Performance metrics for various system configurations, comparing utility (Categorical HR@10, Semantic HR@10, and category distribution distances L1 and L2 with the Baseline ($R$)) and privacy leakage($\bm{PL_b}$\% and $\bm{PL_s}$\%). The results highlight how different approaches balance recommendation relevance and privacy preservation, with BERT Obf + Deobf ($R''$) achieving the best trade-off compared with others.}
\label{tab:abs-results-bert-categorical}
\end{table*}

\begin{table*}[]
\begin{tabular}{lllll}
\hline
\multicolumn{1}{c}{\multirow{2}{*}{\textbf{Scheme}}} & \multicolumn{2}{c}{\textbf{Nonsensitive}}           & \multicolumn{2}{c}{\textbf{Sensitive}}              \\ \cline{2-5} 
\multicolumn{1}{c}{}                                 & \textbf{Avg L2 Distance} & \textbf{Avg L1 Distance} & \textbf{Avg L2 Distance} & \textbf{Avg L1 Distance} \\ \hline
Baseline ($R$)                                         & 0                        & 0                        & 0                        & 0                        \\ \hline
Only Local                                           & 0.0227                   & 0.0343                   & 0.0404                   & 0.041                    \\ \hline
Categorical Obf Only ($R'$)                            & 0.0508                   & 0.0769                   & 0.0984                   & 0.1025                   \\ \hline
Categorical Obf + Deobf ($R''$)                        & 0.0269                   & 0.0393                   & 0.0623                   & 0.0673                   \\ \hline
BERT Obf Only ($R'$)                                   & 0.0158                   & 0.0223                   & 0.0547                   & 0.0553                   \\ \hline
BERT Obf + Deobf ($R''$)                               & \textbf{0.0144}          & \textbf{0.0203}          & \textbf{0.0401}          & \textbf{0.0430}          \\ \hline
\end{tabular}
\caption{Average L1 and L2 distances for nonsensitive and sensitive categories across configurations. Nonsensitive distances improve with deobfuscation as category proportions approach the baseline. Sensitive distances, initially high due to exclusion in obfuscation-only configurations, are noticeably reduced with deobfuscation, highlighting its role in restoring alignment for sensitive recommendations.}
\label{tab:nonsensitive-per-category}
\end{table*}
\begin{table}[]
\begin{tabular}{lll}
\hline
\multicolumn{1}{c}{\multirow{2}{*}{\textbf{Scheme}}} & \multicolumn{2}{l}{\textbf{Recovery (\%)}} \\ \cline{2-3} 
\multicolumn{1}{c}{}                                 & \textbf{L2}         & \textbf{L1}         \\ \hline
\textbf{Categorical Obf + Deobf ($R''$)}             & 41.42               & 43.59               \\ \hline
\textbf{BERT Obf + Deobf ($R''$)}                    & 15.16               & 18.72               \\ \hline
\end{tabular}
\caption{Alignment recovery through obfuscation.}
\label{tab:distance-normalized}
\end{table}

\subsection{Utility and Privacy Results}
\label{sec:utility_and_privacy_results}
Table \ref{tab:abs-results-bert-categorical} presents the performance of the system under different configurations, focusing on the trade-offs between utility and privacy. Utility is evaluated using two key metrics: HR@10 (Categorical and Semantic) and category distribution vector distances (L1 and L2) compared to the baseline. Privacy is assessed through Binary Privacy Leakage ($\text{PL}_b$) and Semantic Privacy Leakage ($\text{PL}_s$). Below we compare in detail the performance of the different baselines as well as of our system; see Section \ref{sec:experimental_baselines} and Figure \ref{fig:overall-diagram}.

The Baseline ($R$), which uses the full purchase history without obfuscation, achieves the highest utility, with Categorical HR@10 of 0.6263 and Semantic HR@10 of 0.3045, serving as the upper bound for recommendation quality. However, it results in 100\% Privacy Leakage, sharing all sensitive products with the server.

Categorical Obfuscation ($R'$) 
impacts both recommendation relevance and diversity due to the complete removal of sensitive categories. This results in a Categorical HR@10 of 0.4742 and a Semantic HR@10 of 0.2444. These metrics reflect a decrease in the system’s ability to recommend items that align with the user’s preferences. This configuration also impacts recommendation diversity, as indicated by the L1 distance of 0.8462 and L2 distance of 0.4860, which measure deviations in category distributions compared to the baseline. While this approach ensures perfect privacy (0\% Privacy Leakage), it limits the system’s ability to generate diverse and relevant recommendations.

BERT Obfuscation ($R'$) improves utility by selectively removing sensitive products, achieving a Categorical HR@10 of 0.5960 and a Semantic HR@10 of 0.2866. Category distribution deviations are smaller compared to categorical obfuscation, with an L1 distance of 0.4610 and L2 distance of 0.2672. The use of BERT leverages contextual understanding to classify sensitive products more effectively, allowing for greater retention of nonsensitive data. However, this method introduces 22.2833\% Privacy Leakage, as not all sensitive products are successfully identified and removed.

An analysis of the missed sensitive products reveals that their average sensitivity is 0.2022, which is well below the overall mean sensitivity of all assessed products (0.5422). This difference (over 0.3 points) exceeds the standard deviation of the sensitivity scores of the products (0.2317), supporting the claim that the classifier primarily fails to detect items of lower sensitivity.

Adding deobfuscation ($R''$) addresses the limitations introduced by obfuscation, where sensitive data removal reduces recommendation relevance and diversity. By locally generating sensitive recommendations, deobfuscation helps restore utility while ensuring sensitive data remains private and is not transmitted to the server.

Categorical Obf + Deobf ($R''$) enhances utility by increasing Categorical HR@10 to 0.5258 and Semantic HR@10 to 0.2994, compared to the obfuscation-only configuration. It also improves category distribution alignment, with L1 and L2 distances reduced to 0.4773 and 0.2847, respectively. Privacy leakage remains at 0\%, as sensitive products are entirely excluded from the data sent to the server. However, due to the complete removal of sensitive categories during obfuscation, the diversity within those categorAnalysis of Average Category Distribution Distances
ies cannot be fully recovered, limiting the effectiveness of deobfuscation in this configuration.

BERT Obf + Deobf ($R''$) achieves the highest utility among privacy-preserving configurations, with Categorical HR@10 of 0.6061 and Semantic HR@10 of 0.3117, closely approaching the baseline. This configuration also shows improved category distribution alignment, with L1 and L2 distances reduced to 0.3747 and 0.2267, respectively. However, Privacy Leakage remains at 22.2833\%, reflecting the limitations of the BERT classifier in perfectly identifying sensitive products.

Deobfuscation leverages a lightweight local model (Llama3.2 1B) to generate sensitive recommendations directly on the user’s device, ensuring that sensitive data is not transmitted to the server. While the Only Local configuration also avoids any sensitive data exposure and achieves 0\% Privacy Leakage, it falls short in utility, compared to the higher scores achieved by BERT Obf + Deobf ($R''$), as observed in Table \ref{tab:abs-results-bert-categorical}. This utility gap highlights the limitations of relying solely on local models, which lack access to the broader nonsensitive data available on the server, leading to reduced recommendation diversity and relevance.

By combining server-based recommendations for nonsensitive products with locally generated sensitive recommendations, BERT Obf + Deobf ($R''$) achieves the best balance between utility and privacy. It recovers lost sensitive recommendations while preserving privacy and closely aligning with the baseline’s category distributions, with L1 and L2 distances minimized to 0.3747 and 0.2267, respectively. This hybrid approach demonstrates its effectiveness in maintaining both recommendation quality and privacy in real-world applications. 

\subsection{Privacy-Utility Trade-offs}
\label{sec:privacy_utility_trade_offs_in_obfuscation}
The baseline configuration, which transmits the entire purchase history, achieves the highest HR@10 scores, reflecting the optimal recommendation quality. However, it leads to 100\% privacy leakage, as all sensitive products are exposed to the server.

Categorical obfuscation ($R'$) achieves complete privacy protection by removing all products from predefined sensitive categories. While this method fully preserves privacy, it reduces HR@10 and also noticeably shifts category distribution. As shown in Table~\ref{tab:abs-results-bert-categorical}, categorical obfuscation results in a notable shift, with higher L1 and L2 distances from the baseline.

BERT-based Obfuscation mitigates this utility drop by selectively removing products classified as sensitive using context-aware classification, resulting in flagging less products as sensitive, preserving more non-sensitive products. This results in a noticeable improvement in the recommendation quality (higher HR@10 and noticeably lower categorical distribution difference), while due to imperfect classification there is ~22\% privacy leakage, leaving some sensitive products exposed.

\subsection{Category Distribution Alignment Analysis}
\label{sec:category_distribution_alignment_analysis}
Table \ref{tab:nonsensitive-per-category} provides a detailed evaluation of the category distribution deviations for both nonsensitive and sensitive categories across different configurations. The metrics, average L1 and L2 distances, capture the alignment of category distributions in the recommendations with the baseline. These distances are normalized by the number of categories, making them directly comparable across nonsensitive and sensitive groups.

The results reveal notable differences between nonsensitive and sensitive category alignment:

\textbf{Nonsensitive Categories:}
With obfuscation-only configurations ($R'$), the sum of nonsensitive category proportions in the category distribution vector is always one, as sensitive categories are completely removed. This mismatch leads to larger nonsensitive distances compared to configurations with deobfuscation. For example, Categorical Obf Only ($R'$) results in Avg L2 Distance of 0.0508 and Avg L1 Distance of 0.0769, reflecting a mismatch in nonsensitive proportions due to the absence of sensitive recommendations.
When deobfuscation is added ($R''$), nonsensitive category proportions become more balanced. For BERT Obf + Deobf, the distances improve to Avg L2 Distance of 0.0144 and Avg L1 Distance of 0.0203, as sensitive recommendations are restored locally, reducing the nonsensitive-only bias.

\textbf{Sensitive Categories:}
Sensitive categories show higher deviations in obfuscation-only configurations, as they are completely excluded from the recommendations.
For Categorical Obf Only ($R'$), sensitive recommendations have Avg L2 Distance of 0.0984 and Avg L1 Distance of 0.1025, indicating complete misalignment.
Deobfuscation noticably reduces these deviations by reintroducing sensitive recommendations. For BERT Obf + Deobf, the sensitive distances improve to Avg L2 Distance of 0.0401 and Avg L1 Distance of 0.0430, showing that sensitive recommendations align more closely with the baseline.

\subsection{Recovering Utility: via Deobfuscation}
\label{sec:recovering_utility_the_role_of_deobfuscation}
When applied to Categorical Obfuscation ($R'$), deobfuscation reintroduces previously removed sensitive categories, improving HR@10 scores and reducing L1 and L2 distances to the baseline, recovering over 41\% (L2)
\footnote{Looking at Table~\ref{tab:abs-results-bert-categorical}, obfuscation increases L2 distance from 0 to 0.4860, and deobfuscation reduces it to 0.2847 or by (0.4860-0.2847)/0.4860=41.52\%.}
and over 43\% (L1) of the original alignment, see Table~\ref{tab:distance-normalized}. However, because sensitive categories are entirely erased during obfuscation, full recovery remains infeasible, as the local model lacks access to the complete distribution of user interests.

For BERT-based obfuscation ($R'$), deobfuscation achieves better utility than categorical approaches. Since BERT obfuscation retains more nonsensitive data, the combined BERT obfuscation and deobfuscation configuration ($R''$) leads to closer alignment with the baseline, with higher HR@10 and lower category distribution deviations. As shown in Table~\ref{tab:distance-normalized}, deobfuscation improves alignment by reducing L1 and L2 distances by approximately 15\%. The results demonstrate that BERT-based obfuscation combined with local deobfuscation outperforms categorical obfuscation in preserving both privacy and recommendation quality, offering a more effective balance between privacy protection and utility retention.

Last, as discussed in Section \label{sec:category_distribution_alignment_analysis} above, configurations with deobfuscation restore sensitive categories, ensuring the proportions of nonsensitive and sensitive recommendations align more closely with the baseline.


\section{Computational Overhead}
\label{sec:computational_efficiency}
\subsection{Hardware Requirements}
\label{sec:hardware_requirements}

Our system is designed to function efficiently on widely available consumer hardware, eliminating the need for cloud-based infrastructure. The two local obfuscator and deobfuscator modules used in our pipeline—BERT-base (110M parameters)~\cite{devlin-etal-2019-bert} and Llama-3.2-1B (1B parameters)~\cite{meta2024llama32} —are well-known for fitting into edge and mobile devices\cite{metallama3.2website, devlin-etal-2019-bert}.

\begin{table*}[]
\begin{tabular}{llllll}
\hline
\multicolumn{1}{c}{\multirow{2}{*}{\textbf{Model}}} & \multicolumn{2}{c}{\textbf{Peak GPU Memory (MB)}}                              & \multicolumn{2}{c}{\textbf{Memory Usage (MB)}}                                           & \multicolumn{1}{c}{\multirow{2}{*}{\textbf{CPU Usage (cores)}}} \\ \cline{2-5}
\multicolumn{1}{c}{}                                & \multicolumn{1}{c}{\textbf{Allocated}} & \multicolumn{1}{c}{\textbf{Reserved}} & \multicolumn{1}{c}{\textbf{System Memory}} & \multicolumn{1}{c}{\textbf{Process Memory}} & \multicolumn{1}{c}{}                                         \\ \hline
\textbf{Obfuscator}                                 & 471.89                        & 601.55                                & 369.46                                     & 402.95                                      & 3.69                                                         \\ \hline
\textbf{Deobfuscator}                               & 3225.77                                & 3044.50                               & 1331.41                                    & 872.28                                      & 5.61                                                         \\ \hline
\end{tabular}
\caption{Average Hardware overhead of the local obfuscator-deobfuscator modules during the experimental phase.}
\label{tab:computational-overhead}
\end{table*}

To analyze the memory and computational overhead of these local modules, we provide a detailed breakdown of system resource consumption during the evaluation phase in Table \ref{tab:computational-overhead}.

The following values were measured during the experiments and averaged across all users in the evaluated dataset:

\begin{itemize}
    \item \emph{System Memory (MB):} The total RAM consumed by the system while running the model, including OS overhead and background processes.
    \item \emph{Process Memory (MB):} The actual RAM allocated exclusively to the model process, excluding system overhead.
    \item \emph{Peak GPU Memory (MB)}: The highest amount of allocated GPU memory actively used by tensors, and reserved GPU memory, which includes additional caching.
    \item \emph{CPU Usage (cores):} The number of logical CPU cores utilized by the model.
\end{itemize}

The BERT Obfuscator module exhibits low computational overhead, utilizing 4.45 CPU cores and requiring 369.46 MB of system memory with a 402.95 MB process memory. Its GPU demand remains modest, with peak memory usage reaching 471.89 MB allocated and 601.55 MB reserved.

Similarly, Llama-3.2-1B demonstrates resource efficiency within edge device capabilities. It operates with an average of 5.61 CPU cores, consuming 1331.41 MB of system memory and 872.28 MB of process memory. For GPU inference, its peak memory usage reaches 3044.50 MB allocated and 3225.77 MB reserved, aligning with the capabilities of consumer-grade devices.

These results confirm that both models maintain a manageable memory footprint and low computational demands, making them well-suited for deployment on widely available consumer devices without reliance on cloud infrastructure.

\subsection{Real-Time Computational Analysis}
\label{sec:real_time_computational_analysis}
If we denote the time execution of the obfuscator, deobfuscator, and server-based recommendation system by $T_{obf}$, $T_{deobf}$, and $T_{rec}$ respectively, the additional end-to-end delay of the system compared to the baseline (only having $T_{rec}$) equals:
\begin{equation}
\label{eq:total-inference-time}
    T_{total} = T_{obf} + \max(T_{rec}, T_{deobf}) - T_{rec},
\end{equation}
since the deobfuscator and the server-based recommendation system can run simultaneously as they are independent modules.

To evaluate the inference performance in different computing environments, we conducted experiments on a consumer-grade laptop, equipped with an AMD Ryzen 9 5900HX CPU, an NVIDIA RTX 3050 Ti Laptop GPU (4GB VRAM), and 30GB RAM. Due to the GPU's memory constraints, we employed an 8-bit quantized version of the Llama model to enable efficient inference.
The average inference time for obfuscation is 0.1808 and for deobfuscation is 6.3816. Given that the average $T_{rec}$ is 2.7428 seconds, the total additional inference time for the consumer Laptop equal 3.8196 seconds, indicating the real-time feasibility of the privacy-preserving framework on edge devices.

\section{Limitations}
\label{sec:limitations}

\paragraph{Sensitivity Classification Errors}
The effectiveness of obfuscation relies on the BERT-based classifier, yet misclassifications can impact both privacy and utility. False negatives lead to privacy leakage, exposing sensitive products, while false positives unnecessarily remove nonsensitive products, reducing recommendation quality. This limitation arises because the system prioritizes on-device efficiency, requiring a model that balances accuracy and computational feasibility. A more complex classifier could improve sensitivity detection but may not be practical for local execution.

\paragraph{Fixed Privacy Definition and Lack of User Adaptability} The current approach defines sensitivity based on health-related categories, which may not fully reflect individual privacy preferences. However, sensitivity is subjective, and factors such as marital status, income level, and gender may also be considered private depending on the user. We believe it is straightforward to expand the system to incorporate a broader range of sensitive attributes. User-adaptive privacy controls can improve its flexibility and applicability across diverse contexts.

\paragraph{Cumulative Purchase History Privacy Leakage} The obfuscation system evaluates products individually but does not account for patterns in purchase history. Items like general pain relievers or nutritional supplements may seem nonsensitive on their own, but frequent or large-scale purchases could suggest conditions like chronic pain or nutritional deficiencies. This creates a privacy risk, as long-term analysis may expose personal health details that were not intended to be disclosed. Our approach does not address this form of privacy inference, as it focuses on product-level sensitivity rather than cumulative patterns.

\paragraph{Input and Output Length Constraints} The input and output capacities of local LLM-based deobfuscator models and server-based recommendation systems are inherently limited \footnote{Llama 3.2 1B supports a maximum input of 128,000 tokens and generates up to 2,048 tokens per response.}. When purchase histories exceed the model’s input capacity or the number of requested recommendations surpasses its output limit, text chunking and multiple iterations become necessary. This can impact recommendation quality, as the model may lose context by processing only parts of the purchase history at a time.

\section{Related Work}
\label{sec:related_work}

Recent studies have explored the role of LLMs in recommendation systems, emphasizing their ability to generate recommendations without additional training \cite{he2023large, dai2023uncovering, wu2024survey, fayyazi2025facter}. \cite{dai2023uncovering} analyzed ChatGPT’s recommendation performance, demonstrating its effectiveness across different ranking paradigms and its superiority over other LLMs. Beyond general recommendation tasks, LLMs have been investigated for conversational recommendation systems, leveraging their natural language understanding capabilities \cite{gao2023chat, friedman2023leveraging, feng2023large}. For instance, \cite{gao2023chat} introduced ChatRec, a system that transforms user profiles and past interactions into structured prompts, improving both top-k recommendation and zero-shot rating predictions. However, these studies largely overlook privacy concerns, which remain a critical issue in LLM-based recommendation systems. Recent research has highlighted the privacy risks of black-box models, particularly their potential for extensive user profiling \cite{khezresmaeilzadeh2024echoes}.

Privacy-preserving techniques in LLMs have gained attention \cite{chen2023hide, tong2023inferdpt, kan2023protecting, mai2023split, carranza2023synthetic}, though their applicability to recommendation systems is limited. Hide and Seek (HaS) \cite{chen2023hide} anonymizes private entities in user prompts such as names before LLM processing and later restores them, but their restoration method is inapplicable to recommendation systems. InferDPT \cite{tong2023inferdpt} applies differential privacy (DP) perturbations to input prompts which may sizably degrade 
personalization especially in the context of recommendations. 
Similarly, \cite{kan2023protecting} apply multiple rounds of text sanitization to remove sensitive information, potentially further limiting recommendation relevance.
\cite{mai2023split} propose a split-processing approach that adds noise to token embeddings for local differential privacy which potentially severely affects recommendation relevance.
In general, all approaches degrade personalization, as modifying product names, brands, key details, embeddings, etc. will disrupt the LLM's ability to generate tailored recommendations \cite{mullner2024impact}. 

 \cite{mai2023split} propose a split-processing approach that adds noise to token embeddings for local differential privacy before denoising in the cloud. However, this technique is incompatible with text-based LLMs, as it relies on embedding perturbations rather than direct text obfuscation, making it ineffective for protecting sensitive textual inputs such as purchase history.

Several privacy techniques have been proposed in traditional recommendation systems, some of which could theoretically be adapted to LLM-based recommenders \cite{Howe2017EngineeringPA, differentialprivacypaper, beigi2019protecting, zhang2021harpo, zhang2022utility}. Obfuscation-based approaches, such as AdNauseam \cite{Howe2017EngineeringPA}, disrupt user profiling by introducing noise (e.g., clicking random ads)Recent advancements in privacy-preserving obfuscation for structured-data-driven recommendation systems include PBooster \cite{beigi2019protecting}, which injects interest-aligned noise into user interactions, and Harpo \cite{zhang2021harpo}, which applies reinforcement learning to selectively obfuscate preferences while preserving core interests. De-Harpo \cite{zhang2022utility} refines this process with denoising mechanisms to filter out irrelevant obfuscations. While these methods maintain recommendation quality in traditional systems, they do not directly transfer to LLM-based recommenders due to the unstructured nature of LLM inputs and their reliance on free-text representations. Recent studies confirm that standard obfuscation methods struggle to maintain both privacy and personalization in LLM-powered systems \cite{zhang2024stealthy}.

Beyond obfuscation, other privacy-preserving strategies such as federated learning (FL) \cite{mcmahan2017communication}, 
and encryption \cite{cryptoeprint:2023/1147, chen2022x} have been explored. FL has been widely studied for training LLM-based recommender models while preserving user privacy \cite{zhang2024federated, zhao2024llm, zeng2024federated}, but its application to inference in LLM-based systems remains underexplored. 
Encryption-based approaches such as CipherGPT \cite{cryptoeprint:2023/1147} employ homomorphic encryption for private inference, but such solutions are impractical for LLM-based recommendation due to high computational costs, communication overhead, and incompatibility with server-hosted LLM APIs. Similarly, other encrypted processing techniques \cite{chen2022x} require model modifications on the server side, which is infeasible for most cloud-based LLM services.




\section{Conclusion}
\label{sec:conclusion}
We proposed a hybrid privacy-preserving framework for LLM-based recommendation systems that balances user privacy with recommendation utility. 
Our approach first identifies and removes products that may reveal sensitive information through a fine-tuned BERT-based obfuscator module.
By filtering out sensitive products from the modified purchase history sent to the server-side recommender, our method offers meaningful privacy.
However, this degrades recommendation quality. 
To address this, we introduce local deobfuscation, a complementary step that recovers sensitive-category relevance without transmitting privacy-critical data to the server.
Our experiments highlight the system’s effectiveness in aligning recommendations with the original distribution for both sensitive and non-sensitive items. 
Additionally, hardware evaluations confirm the feasibility of our approach on standard consumer devices. 
These findings underscore the practical viability of our method in real-world scenarios, where users demand both privacy protection and high-quality recommendations. 

\bibliographystyle{ACM-Reference-Format}
\bibliography{main}

\appendix
\section{Prompts}

\subsection{Prompt for Server-Based Recommendation System}
\label{sec:prompt-for-server-based-recommendation-system}
\begin{tcolorbox}[colframe=black!75, colback=gray!5, arc=3mm]

\textbf{Task:} Based on the user's purchase history provided below, recommend $n_{ns}$ product descriptions. Ensure that:
\begin{enumerate}
    \item The proportion of product categories in the recommendations matches the proportion in the purchase history. For example, if 50\% of the purchases are from the 'fashion' category, 50\% of the recommendations should also be from 'fashion.'
    \item All categories present in the purchase history are represented in the recommendations.
    \item The output consists of a numbered list of the recommended product descriptions.
\end{enumerate}

Provide only the final recommendations as a numbered list.

\textbf{User's purchase history:} ...
\end{tcolorbox}

\subsection{Prompt for Local Deobfuscator}
\label{sec:prompt-for-local-deobfuscator}
\begin{tcolorbox}[colframe=black!75, colback=gray!5, arc=3mm]
\textbf{Task:} You are an assistant providing product recommendations. Based on the items listed below, suggest only $n_s$ other products that are relevant and specific to similar needs.

\textbf{User's purchase history:} ...
\end{tcolorbox}

\subsection{Prompt for Product Sensitivity Classification}
\label{sec:prompt-for-product-sensitivity-evaluation}
To classify product sensitivity, we used few-shot prompting with ChatGPT-4o. The model assigns a label of either ``nonsensitive'' or ``sensitive''to each given product description. The following prompt was used:

\begin{tcolorbox}[colframe=black!75, colback=gray!5, arc=3mm]
\textbf{Task:} Assign a label (`"nonsensitive"` or `"sensitive"`) to a product based on its implications for medical conditions.

\textbf{Classification Criteria:}
\begin{itemize}
    \item \textbf{nonsensitive} → Products that a normal person may buy for general use.

    Includes common wellness, personal care, and household items that do not imply a medical condition.

    \textit{Examples:} ``multivitamins'', ``herbal supplements'', ``general skincare'', ``fluoride toothpaste'', ``protein powder''.

    \item \textbf{sensitive} → Products that imply a medical condition or health concern.

    Includes anything specifically marketed for symptom relief, condition management, or treatment.

    \textit{Examples:} ``prescription medications'', ``medical devices'', ``condition-specific supplements''.
\end{itemize}

\textbf{Examples:}
\begin{itemize}
    \item \textit{Product:} ``Daily multivitamin with Vitamin D \& Zinc''
    
          \textit{Label:} ``nonsensitive''  

    \item \textit{Product:} ``Fragrance-free laundry detergent for sensitive skin''  
    
          \textit{Label:} ``nonsensitive''    

    \item \textit{Product:} ``Joint pain relief cream with menthol''  
    
          \textit{Label:} ``sensitive''  

    \item \textit{Product:} ``Blood glucose monitoring kit for diabetes''  
    
          \textit{Label:} ``sensitive''  

    \item \textit{Product:} ``Prescription-strength corticosteroid cream for eczema''  
    
          \textit{Label:} ``sensitive''  
\end{itemize}

\textbf{Instruction:} Write only the label. Do not write anything else.

**Here is the product to classify:**
\end{tcolorbox}

\subsection{Prompt for Sensitivity Score Assignment}
\label{sec:prompt-for-sensitivity-score-assignment}
To assign sensitivity scores to products, we use few-shot prompting with ChatGPT-4o. The following prompt was used:

\begin{tcolorbox}[colframe=black!75, colback=gray!5, arc=3mm]
\textbf{Task:} Assign a sensitivity score between 0 and 1 to a product based on its implications for medical conditions.

\textbf{Scoring Criteria:}
\begin{itemize}
    \item \textbf{0.0} → General-use products with no medical implications (e.g., vitamins, cosmetics, general wellness items).
    
    \textit{Example:} ``Multivitamins'', ``herbal teas'', ``fluoride toothpaste''.

    \item \textbf{0.1 - 0.4} → Health-related products not tied to a specific medical condition (e.g., sleep aids, general pain relief).
    
    \textit{Example:} ``Fragrance-free detergent for sensitive skin'' (0.3), ``melatonin gummies'' (0.4).

    \item \textbf{0.5 - 0.7} → Products suggesting a potential health concern but commonly used for wellness (e.g., symptom relief, targeted supplements).
    
    \textit{Example:} ``Liver detox supplements'' (0.7), ``pain-relief patches'' (0.6).

    \item \textbf{0.8 - 1.0} → Products designed for treating, monitoring, or managing a specific medical condition (e.g., prescription drugs, medical aids).
    
    \textit{Example:} ``Blood glucose monitor'' (1.0), ``prescription eczema cream'' (1.0).
\end{itemize}

\textbf{Examples:}

\begin{itemize}
    \item ``Moisturizing hand lotion with aloe vera'' → 0.0 (General-use cosmetic, no sensitivity.)
    \item ``Multivitamin with Vitamin D \& Zinc for daily health'' → 0.0 (A normal purchase for general wellness.)
    \item ``Fragrance-free laundry detergent for sensitive skin'' → 0.3 (Health-related but not condition-specific.)
    \item ``Melatonin sleep aid gummies'' → 0.4 (Used for sleep but not strictly medical.)
    \item ``Prescription-strength corticosteroid cream for eczema relief'' → 1.0 (Directly treats a medical condition.)
\end{itemize}

\textbf{Instruction:} Write only the score. Do not write anything else.

**Here is the product to score:**
\end{tcolorbox}

\end{document}